\documentclass[preprint,11pt]{elsarticle}

\usepackage[T1]{fontenc}
\usepackage{amsmath,amssymb,amsthm}
\usepackage{booktabs}
\usepackage{graphicx}
\usepackage{multirow}
\usepackage{url}
\usepackage{xcolor}

\biboptions{authoryear}
\journal{European Journal of Operational Research}

\newif\ifchecks
\checksfalse
\newcommand{\vcheck}[1]{\ifchecks\textcolor{red}{\footnotesize\textsf{[verify: #1]}}\fi}

\begin{document}

\begin{frontmatter}

\title{How exceptional was the Big Three era?\\ Extremes and
persistence in men's professional tennis}

\author[ie]{Manuele Leonelli}
\ead{manuele.leonelli@ie.edu}
\affiliation[ie]{organization={School of Science and Technology, IE University},
            city={Madrid},
            country={Spain}}

\begin{abstract}
Three players won $66$ of the $81$ Grand Slam titles contested between 2003 and 2023, and their era is widely held to be the most dominant in the history of tennis. Assessing it means comparing players who never met, so that every comparison passes through the opponents each did face. We measure dominance by how far a player stands above the field of his own day, how many stand there at once, and how long they remain, the last through averages over fixed windows rather than through unbroken runs, which are unstable when the strengths beneath them are themselves estimated. Strengths come from a Bayesian dynamic Bradley-Terry state space model, and dominance is characterized through exceedances of a high threshold. Applied to $197{,}926$ matches played between 1968 and 2025, only Djokovic separates from the field on peak strength, Federer and Nadal being indistinguishable from Borg, McEnroe and Lendl. What distinguishes the recent period is not how high its best players stood but how long three of them stood there together. At its height three players held a top-three position in at least four fifths of all weeks, where in the strongest comparable decade of the past those positions were shared among four. The sharper anomaly lies between: for twelve years from 1990 the upper tail was not merely unoccupied but compressed.
\end{abstract}


\end{frontmatter}

\section{Introduction}
 
Between Roger Federer's first Grand Slam title at Wimbledon in 2003 and Novak
Djokovic's twenty-fourth at the US Open in 2023, Federer, Rafael Nadal and
Djokovic won $66$ of the $81$ major titles contested, a share of $81.5$ per
cent, and only ten other players won a major over those two decades
\citep{atpslams}. Between them the three occupied the top of the world ranking
for $947$ weeks, and Djokovic alone accumulated $428$ of them, more than any
player before him \citep{atpbigthree}. The period is routinely described as the
most dominant by any group in any sport, and comparisons are usually drawn with
the rivalry of Bj\"orn Borg and John McEnroe in the early 1980s or with the
years of Pete Sampras and Andre Agassi in the 1990s. This paper asks what
such a comparison requires, and whether the available statistical machinery
can deliver it.
 
A player active in 1980 never meets a player active in 2015, so any
statement ranking the two rests on a chain of indirect comparisons mediated by
the opponents they did face. Dynamic paired-comparison models construct such
chains by converting match outcomes into a latent strength trajectory for each
player, and the technology is mature. Our concern is with what happens next.
The claim that three players dominated
the sport for two decades is not a claim about any one of them. It bundles
together three separate questions: how far the best players stood above the
rest of the tour, how many of them stood there at the same time, and how long
that arrangement lasted. The first asks about the magnitude of an extreme, and
the second about how many extremes coincide; for both, the peaks-over-threshold
apparatus of extreme value theory supplies the standard tools. The third asks
how long a configuration is maintained, and it is the one that usually carries
most of the rhetorical weight. It is also the one whose answer depends most
sharply on how the question is posed, and part of what follows is concerned
with establishing which ways of posing it can be trusted.
 
The difficulty is easiest to see through an example. Fix a level high enough
that only a dominant player clears it, and ask how long three players held it
at once. One kind of answer counts weeks: in what fraction of the weeks between
2003 and 2022 were three players simultaneously above the level, and, given
that three were above it in some week, how likely is that still to be true a
year later? Another measures an unbroken stretch: what was the longest run of
consecutive weeks with three players continuously above it? Both are natural
descriptions of the same phenomenon and both appear in applied work, but they
respond very differently to the smallest disturbance. Suppose that in a single
week of 2012 Federer's strength dips just below the level, on a minor injury,
an early loss, or nothing more than noise in the rating. The counting answers
move by one week in a thousand. The unbroken stretch is halved, a decade of
continuous three-way dominance becoming two spells of five years. Averaging
over the record absorbs an isolated interruption; deciding week by week whether
a spell has ended does not. We report the first kind throughout and avoid the
second. Unbroken stretches are delicate wherever a record must be cut into clusters,
and more so when the strengths beneath them are estimated rather than
observed; the same difficulty arises for any latent state that is filtered
and then thresholded, and sport supplies an unusually complete record on
which to examine it.
 
Answering questions of this kind, about how far above the field, how many at
once and for how long, requires a strength for every player in every week,
carrying honest uncertainty. We estimate these with a Bayesian dynamic Bradley-Terry state space model
\citep{durbinkoopman2012}. Each player is given a latent strength evolving as a
random walk. The same model is later given one component per playing
surface, correlated across them, so that a result on clay revises what we
believe about a player on grass in proportion to how far the surfaces share
information. Rather than a single fitted rating, we take samples from
the posterior over entire strength trajectories and recompute every quantity
within each sample, so that uncertainty about the ratings reaches every number
we report. A player counts as dominant in a week if his strength exceeds a
threshold, and that threshold is set not in rating points but so that, on
average over the whole period, a stated number of players clear it each week.
Eras are then compared at a common rate of exceedance rather than a common
level. This matters because what is comparable across five decades is not a
distance in rating points but how rare it was to stand that far above one's own
contemporaries.
 
Applied to the complete record of men's professional tennis from 1968 to 2025,
the picture that emerges is not quite the conventional one. On peak strength
only Djokovic separates from the field; Federer and Nadal cannot be
distinguished from Borg, McEnroe and Lendl. Nor, on a permissive definition of
dominance, were three players dominant at once more often than in the 1980s.
What is exceptional is duration, and only for the first half of the era: the
probability that three-way dominance survives a full year is $0.44$ between
2002 and 2013, against $0.12$ in the early 1980s and $0.06$ thereafter. The
years in between are the sharper anomaly. From 1990 to 2001 the upper tail was
not merely unoccupied but compressed, and some of the sense that the era which
followed was unprecedented derives from the contrast with the one that
preceded it.

\section{Literature review}
\label{sec:lit}

\subsection{Paired comparisons and dynamic ratings}

The modern treatment of paired-comparison data begins with the model of
\citet{bradley1952}, in which the probability that one player defeats another
is a logistic function of the difference between two latent strengths;
\citet{cattelan2012} surveys the subsequent development, with particular
attention to the dependence structures that arise when comparisons share
players. Extending the static formulation to settings in which strengths evolve
is the central methodological problem in competitive ranking, and two broad
approaches have emerged. The first is recursive and heuristic, exemplified by
the system of \citet{elo1978} and its descendants, in which ratings are updated
after each match by an amount proportional to the discrepancy between the
observed and expected result; comparative assessments of such systems in tennis
are given by \citet{kovalchik2016}. The second treats strength as a latent
stochastic process observed through match outcomes, which places the problem
within the framework of nonlinear state space models. \citet{glickman1999} gave
the canonical treatment, approximating the intractable filtering recursions by
Gaussian updates and thereby producing ratings accompanied by uncertainty
measures. \citet{gorgi2019} pursue a
related high-dimensional dynamic formulation for tennis, and
\citet{klaassenmagnus2001} provide the point-level evidence that motivates
treating match outcomes as informative but noisy realizations of an underlying
strength difference.

This literature is well represented in operational research.
\citet{bakermchale2014} developed a dynamic paired-comparison model with the
explicit aim of comparing tennis players across eras. Their solution, a barycentric rational interpolation of time-varying
strengths combined with empirical Bayes shrinkage, was applied to the women's
tour in \citet{bakermchale2017}. Further contributions in the same tradition
include weighted variants of the Elo recursion \citep{angelini2022} and
network-based measures of competitive standing \citep{arcagni2023}. The question we address is therefore not new. We differ from this work in treating the ratings as objects carrying
uncertainty that must be propagated, which rules out fitted trajectories such
as the barycentric interpolants of \citet{bakermchale2014}, since a single
deterministic trajectory per player carries none, and in asking which
downstream summaries survive that propagation.

A parallel literature concerns competitive balance and its evolution.
\citet{berry1999} confronted the era-bridging problem directly, developing a
hierarchical model to compare athletes whose careers do not overlap.
\citet{radicchi2011} constructed a prestige measure from the network of matches
and used it to rank players across the professional era, while
\citet{breznik2025} asked whether particular periods constitute eras of
concentrated dominance, identifying several through cointegration analysis of
ranking trajectories. The latter is the closest antecedent to our substantive
question, and we return to the comparison in Section~\ref{sec:disc}.

\subsection{Extremes in sport}

Characterizing dominance as behaviour in the upper tail connects naturally to
extreme value theory, whose general apparatus is set out in \citet{coles2001}.
The literature applying these ideas to sport has developed almost entirely
within disciplines in which performance is directly measured on a continuous
scale. \citet{einmahlmagnus2008} estimated ultimate records across athletic
events, and the same framework has since been applied to sprinting records in
a formulation permitting athlete-specific tails \citep{einmahlhe2026}.
\citet{stephensontawn2013} address the related problem of ranking performances
through a conceptual population model, and swimming has received careful
treatment in \citet{spearing2021}.

A recurring observation in that literature is that dependence structure is
typically neglected: analyses either assume that repeated performances by the
same athlete are independent, or reduce each athlete to a single best result
\citep{spearingtawn2023}. The framework of \citet{spearingtawn2023} for extremes
of longitudinal data is designed to remedy this and is the closest
methodological antecedent to the present work, although it is developed for
directly measured performances rather than for latent strengths that must
themselves be inferred. That difference matters here, though not only for the reason one might expect.
Estimation error does interact with the extremal summaries, but the summaries
themselves differ in how much disturbance they can absorb, and the second point
is the more fundamental.

The question of whether several extremes occur together has been formalized in
the spatial setting by \citet{dombry2018}, who introduced extremal concurrence
as the event that the maximum over time is achieved simultaneously at several
locations. Related notions appear in the climate literature under the heading of
compound and concurrent events \citep{zscheischler2020}, whose spatially
compounding case is the closest to our object, with players playing the role
of locations. It differs from the concurrence probability
of \citet{dombry2018} in a respect made explicit in Section~\ref{sec:func}: we
are concerned with sustained joint occupation of the upper tail rather than with
the coincidence in time of individual maxima. The two need not agree, since
players may hold the tail together for years while peaking at different moments,
and we estimate the former rather than the latter.

\subsection{Two-step inference}

Estimating strengths and then summarizing them is a two-step procedure, and
each step has a literature on what can go wrong. The first concerns the
summary. A statistical functional is qualitatively robust when small
perturbations of the underlying distribution produce small changes in its value
\citep{hampel1971}, and the longest run of exceedances is not: moving one period across the threshold can halve it. This
is why cluster-based summaries are known to be delicate even when the process
is fully observed, as the literature on estimating the extremal index, whose
reciprocal is the mean cluster size, attests
\citep{ferrosegers2003,northrop2015}.
 
The second concerns the inputs, which here are estimated rather than measured.
The difficulty is long recognized for generated regressors
\citep{pagan1984} and has become more pressing wherever fitted or predicted
values are passed downstream as though they were data
\citep{leonellismith2015,angelopoulos2023}. The remedies proposed in that literature do not transfer here, since the
difficulty is not one of variance but of bias and the direction of the bias
reverses between estimators. The two difficulties compound: estimation error is exactly the sort of perturbation to which a non-robust summary is sensitive, and a rating scheme supplies it systematically rather than at random. Section~\ref{sec:sim}
measures what this costs.

\section{Dominance, exceedances and their summaries}
\label{sec:func}

This section defines what we mean by dominance and sets out the summaries we
shall report. Nothing in it depends on the model of Section~\ref{sec:model}. We
require only some procedure, which may be a recursive rating system, the
posterior mean of a statistical model, or a single draw from a posterior, that
assigns to each player $i$ and period $t$ a strength $\theta_{i,t}$.

\subsection{Relative strength}

Absolute strength is not comparable across periods, because the reference level
implied by any rating scheme drifts as the population of players turns over. We
therefore define the relative strength of player $i$ in period $t$ as
\begin{equation}
r_{i,t} = \theta_{i,t} - \bar\theta_t^{(10:100)},
\label{eq:rel}
\end{equation}
where $\bar\theta_t^{(10:100)}$ is the mean strength of the players ranked tenth
to hundredth among those active in period $t$. Excluding the top nine prevents
the reference level from being contaminated by the very players whose dominance
is under study, while excluding the long tail of marginal participants prevents
it from tracking changes in the size and composition of the professional
population. Dominance is thus defined relative to the contemporaneous field
rather than on an absolute scale, which is what makes cross-era comparison
meaningful.

Relative strengths are reported throughout on the conventional rating scale,
obtained by multiplying the latent difference by $400/\log 10$ so that $400$
points correspond to ten-to-one odds between two players. A relative strength
of $300$ therefore means that the player would be expected to win roughly
$85\%$ of matches against a player at the level of the reference band. These
are not Association of Tennis Professionals (ATP) ranking points, which
accumulate results over a season and measure
something different. The choice of reference band is a judgement, and
Section~S4 of the supplementary material reports how the results change when it is widened or
narrowed.

\subsection{Dominance as a peaks-over-threshold problem}

Let $u$ be a threshold and define the exceedance count
\begin{equation}
N_t(u) = \sum_i \mathbf{1}\{r_{i,t}>u\},
\label{eq:count}
\end{equation}
the number of players standing more than $u$ above the field in period $t$. The claim that a sport was dominated by three players amounts to the claim that $N_t(u)\geq 3$ for a high $u$ and for a long time, and everything we report is a summary of the process $\{N_t(u)\}$ or of the exceedances themselves.

This is the peaks-over-threshold construction \citep{coles2001}, with two
departures from a standard application. The first is that $r_{i,t}$ is latent,
which is the subject of the paper. The second is that exceedances are strongly
dependent both across players within a period, since $r_{i,t}$ is defined
relative to a common field, and across periods within a player, since strength
evolves smoothly. It is the second kind of dependence that the summaries below
are designed to describe. The magnitudes of the exceedances are reported in
Section~\ref{sec:results} through order statistics and career peaks rather
than through a fitted generalized Pareto tail, since the claims under
examination concern how many players clear the threshold and for how long.

One choice has to be made before any of this can be stated precisely. The
threshold $u$ requires care, because a latent strength is identified only up to
scale, so a threshold fixed in absolute units is not comparable across rating
schemes or model specifications. We therefore calibrate $u$ so that the
marginal exceedance rate is fixed by construction: given a target mean count $\bar N$ over an
analysis window of $T$ periods, we set $u$ equal to the $(\bar N T)$-th largest
value of $r_{i,t}$. When strengths are drawn from a posterior, this calibration
is performed separately within each draw. Every comparison we report is
therefore made at a common marginal rate, which removes scale as a confounder
and isolates the distributional question of how exceedances are arranged in
time. The calibration fixes only the mean of $N_t$ over the window; the
probability that three players exceed the threshold at once, and how that
event is arranged in time, remain free to differ between eras, and are what
is compared. We work throughout with two values of $\bar N$, one permissive and one
demanding, since the definition of dominance is a matter of degree and the
comparison between eras turns out to depend on it in an informative way. Each
table states which threshold it uses.

\subsection{What we estimate}

Three summaries of $\{N_t(u)\}$ are reported throughout. The first is how often
the configuration occurs, through the rate $P(N_t\geq c)$ and the expected
count $E[N_t]$. The second is the lag persistence probability
$\pi_k = P(N_{t+k}\geq c \mid N_t\geq c)$, the chance that a configuration in
place now is still in place $k$ periods later. The third is the block
persistence probability
$\beta_w = P(N_s\geq c \text{ for all } s\in[t,t+w))$, the chance that it holds
without interruption throughout a window of $w$ periods. Each is a time average
of a bounded function of a fixed-width window, and between them the profiles
$\{\pi_k\}$ and $\{\beta_w\}$ convey what a claim about the duration of
dominance is usually meant to convey.

The obvious alternative is the longest unbroken run of periods with
$N_t\geq c$, and we avoid it deliberately. Summaries of that kind require the
record to be partitioned into clusters of exceedance, which is delicate even
when the process is observed, as Section~\ref{sec:lit} noted, because the
answer depends on the threshold and on whatever declustering rule separates one
cluster from the next. Our setting adds a second difficulty. Strengths here are
inferred, so the sequence handed to any such summary carries interruptions of
its own, whose direction depends on the estimator: a smooth recursive rating
manufactures continuity and lengthens clusters, while independently perturbed
posterior samples fragment them. Averages over fixed windows are largely
unaffected, since an isolated interruption contributes in proportion to the
inverse record length instead of reorganizing the whole answer.
Section~\ref{sec:sim} quantifies the difference.

\subsection{How the summaries behave}
\label{sec:sim}

To see what this costs in practice we simulate records for which the answer is
known: $250$ players over $400$ periods, strengths
following independent random walks and match outcomes a logistic link, which is
the process specified in Section~\ref{sec:model}, with roughly $70$ matches per
period among $48$ active players. We repeat this $25$ times and compute every
functional three ways on each replicate: from the true strengths, from a
recursive Elo rating treated as observed data, and as the posterior median
under the scheme of Section~\ref{sec:model}. All three are calibrated to a
common marginal exceedance rate, so no comparison is confounded by the choice
of threshold.

\begin{table}[t]
\centering
\small
\caption{Behaviour of each functional across $25$ simulated records from one
process. The first column is the coefficient of variation of the true value
across replicates, computed without estimation. The remaining columns give the
mean and standard deviation (s.d.) across replicates of the ratio of estimate
to truth, so that unity denotes exact recovery, when a recursive rating is
treated as observed data and when the posterior median is used. All estimators
are calibrated to a common marginal exceedance rate, which fixes the
coefficient of variation of the rate functional. \vcheck{script 06 and the truth block of the accompanying code}}
\label{tab:sim}
\begin{tabular}{lccccc}
\toprule
& truth & \multicolumn{2}{c}{rating as observed} & \multicolumn{2}{c}{posterior median} \\
\cmidrule(lr){2-2}\cmidrule(lr){3-4}\cmidrule(lr){5-6}
Functional & c.v. & mean & s.d. & mean & s.d. \\
\midrule
\multicolumn{6}{l}{\textit{Run-based summaries}}\\
Longest run                 & 0.47 & 1.75 & 0.87 & 1.32 & 0.60 \\
Mean run length             & 0.60 & 3.56 & 2.67 & 1.79 & 0.66 \\
\midrule
\multicolumn{6}{l}{\textit{Gap-tolerant runs}}\\
Tolerance $g=4$             & 0.33 & 1.35 & 0.66 & 1.00 & 0.35 \\
Tolerance $g=8$             & 0.25 & 1.25 & 0.47 & 1.00 & 0.28 \\
\midrule
\multicolumn{6}{l}{\textit{Fixed-window averages}}\\
Lag persistence $\pi_1$     & 0.06 & 1.12 & 0.07 & 1.08 & 0.07 \\
Lag persistence $\pi_8$     & 0.23 & 1.41 & 0.41 & 1.15 & 0.29 \\
Rate functional             & 0.01 & 1.00 & 0.01 & 1.00 & 0.01 \\
\bottomrule
\end{tabular}
\end{table}

Table~\ref{tab:sim} orders the summaries twice over. The first column does so
before any estimator is involved: the longest run and the mean run length vary
by roughly half their own size across replicates of one and the same process,
so a single history says little about the runs that process tends to produce,
whereas $\pi_1$ varies by six per cent and the rate functional not at all. The
remaining columns show what estimation adds. Treating ratings as observed
inflates the two run-based summaries by factors of $1.75$ and $3.56$, while
$\pi_1$ and the rate functional are recovered to within a few per cent.
Gap-tolerant runs fall between the two, and a tolerance of eight periods
removes most of the bias at the price of a coarser description.

Two points are worth adding. Posterior sampling alone does not repair a
run-based summary, improving the longest run only from $1.75$ to $1.32$. The
block persistence probabilities, which are levels rather than ratios, are
overstated at every window and most severely at long ones, at $0.132$ against
a truth of $0.069$ for $w=13$, with the posterior roughly halving the error;
this is a reason to read the whole profile rather than any single window.

\subsection{Dominance across playing surfaces}

Tennis is played on surfaces whose demands differ, so a player may dominate on
one and not another, and joint occupation of the upper tail across surfaces is
itself a form of exceptional achievement. Writing $r^{(s)}_i$ for a player's
career peak relative strength on surface $s$ and $U^{(s)}_i$ for its rank
percentile within a reference population, we summarize cross-surface dependence
through the coefficients of \citet{ledford1996},
\begin{equation}
\chi(q) = \frac{P(U^{(s)}>q,\,U^{(s')}>q)}{1-q},
\qquad
\eta(q) = \frac{\log(1-q)}{\log P(\min(U^{(s)},U^{(s')})>q)},
\label{eq:chieta}
\end{equation}
and we report the number of players simultaneously extreme on all three
surfaces, comparing it against the counts implied by independence and by
perfect dependence.

\section{A dynamic Bradley-Terry state space model}
\label{sec:model}

Section~\ref{sec:func} requires a strength $\theta_{i,t}$ for every player and
period, and this section specifies the model that supplies it.

\subsection{Specification}

Because tennis is played on several surfaces, we take the state to be a vector
rather than a scalar. Let
$\boldsymbol\theta_{i,t}=(\theta_{i,t}^{(1)},\ldots,\theta_{i,t}^{(S)})^{\top}$
denote the latent strengths of player $i$ at time $t$, one component per
surface, where $t$ indexes weekly rating periods. We assume
\begin{equation}
\boldsymbol\theta_{i,t} = \boldsymbol\theta_{i,t-1} + \mathbf{w}_{i,t},
\qquad \mathbf{w}_{i,t}\sim N_S(\mathbf{0},\tau^2 \mathbf{R}(\rho)),
\label{eq:state}
\end{equation}
independently across players, with initial condition
$\boldsymbol\theta_{i,t_{0i}}\sim N_S(\mathbf{0},\sigma_0^2\mathbf{R}(\rho))$ at
the period $t_{0i}$ of a player's first appearance. Here $\mathbf{R}(\rho)$ is
an exchangeable correlation matrix with off-diagonal $\rho$, so that a single
parameter governs how far improvement on one surface carries over to the others.
At $\rho=1$ the model collapses to a single common strength, and at $\rho=0$ the
surfaces become separate problems. Two specializations are used below. Every
result other than those of Section~\ref{sec:surf} takes $S=1$, a single
strength per player estimated from all matches irrespective of surface, which
is the appropriate object when the question concerns dominance over the field
as a whole. Section~\ref{sec:surf} takes $S=3$ and estimates $\rho$, in order to
ask whether occupancy of the upper tail transfers between surfaces.

Conditional on strengths, the probability that player $i$ defeats player $j$ in
a match played on surface $s$ over a maximum of $f$ sets is
\begin{equation}
P(i \text{ defeats } j \mid \boldsymbol\theta_{i,t},\boldsymbol\theta_{j,t})
= \big[1+\exp\{-a_f(\theta_{i,t}^{(s)}-\theta_{j,t}^{(s)})\}\big]^{-1}.
\label{eq:obs}
\end{equation}
Men's professional tennis uses two match formats. Matches at the four Grand Slam
tournaments are played as the best of five sets, and almost all other matches as
the best of three. A longer match affords more opportunity for the stronger
player to prevail, so the same strength difference should translate into a
larger win probability, and we allow this through the discrimination parameter
$a_f$. We normalize $a_3=1$, which fixes the scale of the latent strengths, and
estimate $a_5$, which is identified by the relative frequency of upsets across
the two formats. A match thus informs one component of the state directly, and
the remaining components indirectly through the correlation in
\eqref{eq:state}.

\subsection{The posterior}

Write $\boldsymbol\psi=(\tau,\sigma_0,\rho,a_5)$ for the hyperparameters, which
govern the dynamics and the link and are held fixed at the values estimated in
Section~\ref{sec:hyp}. Given $\boldsymbol\psi$ the model has two levels. The
strength trajectories $\boldsymbol\Theta=\{\boldsymbol\theta_{i,t}\}$ follow the
Gaussian Markov process \eqref{eq:state}, which supplies the prior for
$\boldsymbol\Theta$: it encodes the assumption that strengths change slowly and
by small amounts, and it is what allows a player's estimated strength in one
week to be informed by his results in neighbouring weeks. The match outcomes
$\mathbf{y}$ are then conditionally independent Bernoulli variables with
probabilities given by \eqref{eq:obs}. The posterior is
\begin{equation}
p(\boldsymbol\Theta\mid\mathbf{y},\boldsymbol\psi)\propto\prod_{i}\Big[ p(\boldsymbol\theta_{i,t_{0i}})\prod_{t>t_{0i}}
p(\boldsymbol\theta_{i,t}\mid\boldsymbol\theta_{i,t-1})\Big]
\prod_{m} p(y_m\mid \boldsymbol\theta_{i_m,t_m},\boldsymbol\theta_{j_m,t_m}),
\label{eq:posterior}
\end{equation}
where $m$ indexes matches. Every quantity in Section~\ref{sec:func} is a
functional of $\boldsymbol\Theta$, so all inference proceeds by sampling from
\eqref{eq:posterior} and evaluating the functional within each sample.

Two features of \eqref{eq:posterior} prevent this being done directly. The
state process is Gaussian but the likelihood is logistic, so no conjugacy is
available; and because each match involves two players, the posterior does not
factorize across players, so the state is effectively of dimension $SN$, where
$N$ is the number of players. Exact treatment by Markov chain Monte Carlo
remains possible, and we construct such a scheme in Section~S3 of the
supplementary material, but it does not scale to two hundred thousand
matches. We therefore adopt a deterministic approximation, and
the same section shows that it reproduces the exact posterior closely
on a subsample where both can be run.

\subsection{Estimating the hyperparameters}
\label{sec:hyp}

The hyperparameters are treated by empirical Bayes, fixed at the maximizer of
the prequential likelihood \citep{dawid1984},
\begin{equation}
\ell(\boldsymbol\psi)
= \sum_{m} \log p\big(y_m \mid \mathbf{y}_{1:m-1},\boldsymbol\psi\big),
\label{eq:preq}
\end{equation}
restricted to matches in which both players have prior history, since a
prediction involving a debutant carries no information about the dynamics. For
a state space model this decomposition is the likelihood, subject to the
Laplace approximation used in filtering; the predictive probabilities come from
the filter of Section~\ref{sec:approx}, so maximization is a grid search over
$\boldsymbol\psi$ with one forward pass at each point. Fixing $\boldsymbol\psi$
rather than giving it a prior is itself an approximation, and the component
that matters is $\tau$, the weekly innovation standard deviation. We assess it
through the curvature of \eqref{eq:preq} at its maximum and propagate it by
refitting at displaced values, with the result reported in
Section~\ref{sec:results}.

\subsection{Approximating the posterior}
\label{sec:approx}

We approximate \eqref{eq:posterior} by a sequence of local Gaussian
approximations, following the general strategy of \citet{glickman1999}. Two
simplifications are involved. The first is a mean-field step across players:
within a rating period each player's strength is updated with the opponents'
strengths held at their current values, their uncertainty entering only through
an attenuation factor. The second is a Laplace step within each player: the
period posterior is replaced by a Gaussian centred at its mode with curvature
evaluated there. One point of practice matters. The Glicko rating update takes
a single Fisher scoring step towards that mode, which is exact when a player
plays one match per period; it is not enough here, since a player may contest
seven matches in a fortnight, and in our data the single-step version placed
two players of modest achievement among the highest-rated in the sample. We
therefore iterate the step to convergence, at negligible cost, and the artefact
disappears. The Laplace approximation is then reproduced exactly by a Gaussian
pseudo-observation, which puts the model in linear Gaussian state space form
and lets a result on one surface revise the estimated strength on the others
in proportion to $\rho$. The update equations are given in Section~S2 of the
supplementary material.

\subsection{Trajectory sampling}

The functionals of Section~\ref{sec:func} depend on the joint distribution of
the strength process across adjacent periods, not on its pointwise marginals, so
pointwise posterior summaries are inadequate. We therefore sample complete
trajectories by forward-filtering backward-sampling
\citep{carterkohn1994}. For a player observed in periods
$t_1<\cdots<t_K$ with filtered means $\mathbf{m}_k$ and covariances
$\mathbf{P}_k$, we draw
$\boldsymbol\theta_{t_K}\sim N_S(\mathbf{m}_K,\mathbf{P}_K)$ and then
recursively, for $k=K-1,\ldots,1$,
\begin{equation*}
\boldsymbol\theta_{t_k}\mid\boldsymbol\theta_{t_{k+1}} \sim
N_S\!\left(\mathbf{m}_k + \mathbf{J}_k(\boldsymbol\theta_{t_{k+1}}-\mathbf{m}_k),\;
\mathbf{P}_k-\mathbf{J}_k\mathbf{P}_k\right),
\end{equation*}
where $\mathbf{J}_k=\mathbf{P}_k(\mathbf{P}_k+\mathbf{Q}_k)^{-1}$ is the
smoother gain and $\mathbf{Q}_k=(t_{k+1}-t_k)\tau^2\mathbf{R}(\rho)$ the
innovation covariance accumulated across the gap. 

\section{Eras of dominance on the men's tour}
\label{sec:results}

Every quantity below is an estimand of Section~\ref{sec:func} computed from the
strengths of Section~\ref{sec:model}, evaluated within each of $300$ posterior
samples and summarized across them. Results are reported by era, and the division is mechanical: the window is cut
into four blocks of twelve years, 1978--1989, 1990--2001, 2002--2013 and
2014--2025. Boundaries chosen to bracket the careers of the players under
study would make the comparison less than blind, and blocks of unequal length
would give the longer one more opportunity to accumulate whatever is being
counted. It happens
that 1990 nearly coincides with a genuine break, the replacement of the Grand
Prix and World Championship Tennis circuits by the ATP Tour, and that the third
block opens the year
before Federer's first major title, so the last two between them span the whole
period over which the three players overlapped, and the boundary between them
splits that period in half so that the two halves can be compared. Because no division of a continuous record is beyond argument,
Figure~\ref{fig:conc} presents the same quantity as a continuous series on
which no boundary is imposed, and Section~\ref{sec:best}
compares the strongest decade in the record against the strongest decade
disjoint from it, which requires no division at all.

\subsection{Data}
\label{sec:data}

We analyse the $197{,}926$ men's professional singles matches played on the
ATP tour between 1968 and 2025 and recorded in the TML-Database
\citep{tmldatabase}, which takes its results, player identifiers and
tournament information from the official ATP records. Each match supplies its
date, tournament, surface, format and the identities of the two players;
$188{,}771$ of them
involve two players who have both appeared before and so contribute to
estimation. Matches are aggregated into weekly rating periods. The analysis
window runs from 1978 to 2025, because the reference level of \eqref{eq:rel}
rises steeply through the first decade of the open era as the professional
population stabilizes and is flat thereafter, lying between $1900$ and $1935$
on a conventional rating scale in every five-year block from 1978 onwards; it
contains $1{,}924$ periods and a median of $558$ players active in a period,
a player counting as active if he has played within the preceding fifty-two
weeks. The surface analysis uses the $143{,}383$ matches played on hard, clay
and grass, hard including indoor hard courts, and sets aside carpet, the older
indoor surface, which accounted for $17$ per cent of matches before 1990 but
was withdrawn from the tour and last used in 2017; the rank percentiles of
\eqref{eq:chieta} are computed over the $199$ players with at least forty
matches on each of the three surfaces.

\subsection{Model checking}

On the $188{,}771$ matches in which both players have prior history the fitted
model attains a predictive accuracy of $0.675$ and a Brier score of $0.207$, in line
with published benchmarks for rating-based prediction in tennis
\citep{kovalchik2016}. The prequential likelihood is sharply peaked in $\tau$,
whose posterior standard deviation is under two per cent of its value, and
refitting two standard deviations either side leaves every comparison reported
below unchanged to two decimal places.

In common with rating models generally, ours is somewhat overconfident, and
increasingly so for stronger predictions: in the highest probability band,
predictions averaging $0.934$ correspond to an observed frequency of $0.910$.
Recalibration on the logit scale \citep{platt1999}, estimated on odd-numbered
years and evaluated on even-numbered ones, removes it. The estimated intercept
is zero and the slope $0.851$, so the model overstates every strength
difference by a common factor of about $1.18$, and after correction the
reliability diagram is flat to within $0.006$ in every band. Being a global
monotone rescaling, the correction leaves every ranking, every pairwise
comparison and every rate-calibrated exceedance set unchanged, and affects only
the units in which strengths are quoted. All reported strengths incorporate it.
Selected hyperparameters, the reliability diagram and the details of the
calibration are collected in Section~S1 of the supplementary material.

\subsection{The shape of the top of the field}

Before turning to individual players it is useful to see what the top of the
field looked like in each era. Table~\ref{tab:prof} reports, for each rank
position from first to fifth, the era-average of the corresponding order
statistic of relative strength. This is a time average of a bounded function of the configuration at a single
period, and so among the summaries of Section~\ref{sec:func} that are reliably
estimated.

\begin{table}[t]
\centering
\small
\caption{Era-average of the $k$th largest relative strength, in calibrated
rating points above the contemporaneous field, with $95\%$ credible intervals. \vcheck{script 08, top-five profile block}}
\label{tab:prof}
\begin{tabular}{lcccc}
\toprule
Rank & 1978--1989 & 1990--2001 & 2002--2013 & 2014--2025 \\
\midrule
1 & 397 [376, 424] & 264 [246, 284] & 364 [347, 387] & 377 [350, 400] \\
2 & 341 [325, 358] & 223 [210, 237] & 305 [289, 323] & 321 [302, 340] \\
3 & 296 [278, 315] & 198 [187, 208] & 260 [247, 276] & 284 [268, 302] \\
4 & 256 [239, 272] & 179 [170, 188] & 229 [214, 246] & 237 [221, 257] \\
5 & 220 [205, 235] & 165 [157, 174] & 202 [189, 215] & 207 [195, 222] \\
\bottomrule
\end{tabular}
\end{table}

Two features stand out. The first is that the block of the 1980s is above every
later one at every rank. The strongest player of an average week between 1978
and 1989 stood $397$ points above the field, against $364$ and $377$ in the two
blocks spanning the careers of Federer, Nadal and Djokovic, and the ordering is
the same from first place to fifth. Whatever distinguishes the later period, it
is not that the top of the field stood further above the rest. The second is the depth of the gap between those blocks and the one separating
them. The strongest player of an average week between 1990 and 2001 stood $264$
points above the field, and the fifth-ranked player $165$, a level exceeded by
the eighth or ninth ranked player in the neighbouring blocks. The whole upper
region of the distribution was compressed, not merely its extreme. The
credible intervals of the two neighbouring blocks lie entirely above those of
1990--2001 at every rank, so the compression is not an artefact of rating
uncertainty.

\begin{figure}[t]
\centering
\includegraphics[width=0.7\textwidth]{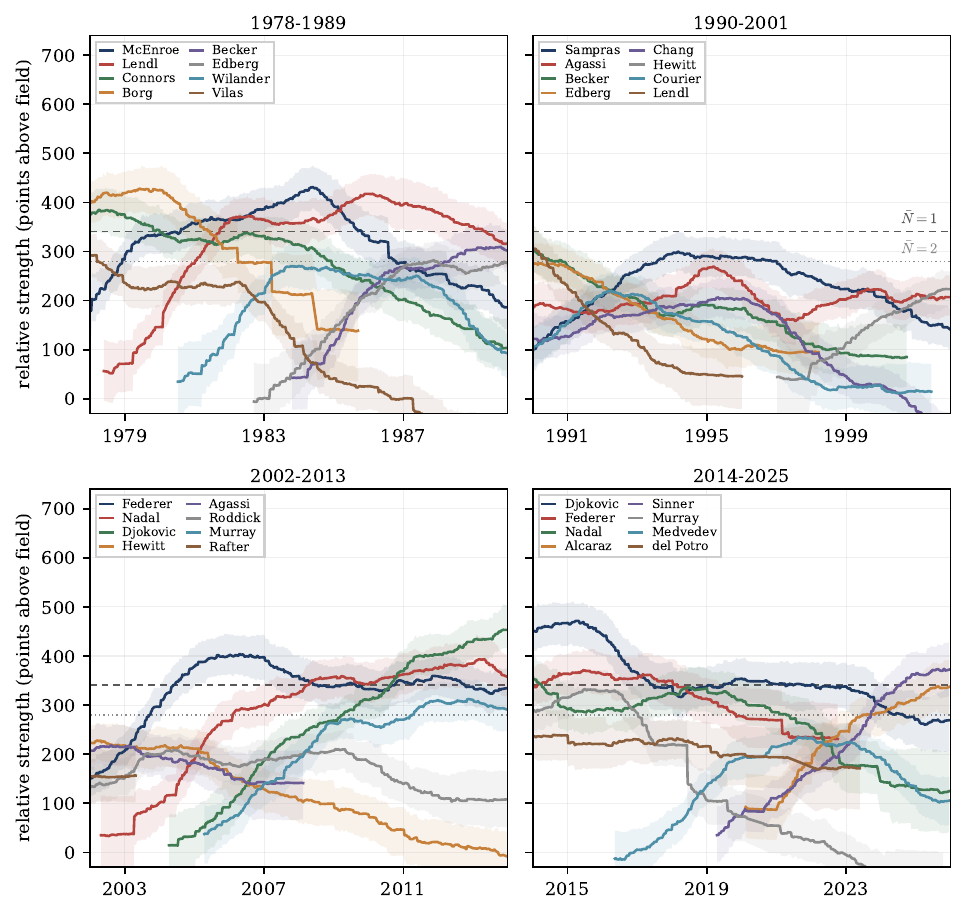}
\caption{Posterior relative strength of the eight players who most often occupied
the top three positions in each block, with $80\%$ credible bands. Horizontal
lines mark the two rate-calibrated thresholds, dashed for $\bar N=1$ at $341$
points and dotted for $\bar N=2$ at $280$ points. All panels share a vertical
scale and span twelve years. \vcheck{scripts 09 then 12, fig8}}
\label{fig:traj}
\end{figure}

Figure~\ref{fig:traj} shows the underlying trajectories, giving for each block
the posterior relative strength of the eight players who most often occupied
the top three, together with the two rate-calibrated thresholds. It makes the same
comparison visible player by player. In the first panel four players cross the
stricter threshold but do so in succession rather than together: Borg is above
it from 1978 until 1981, Connors and McEnroe alternate through the early
1980s, and Lendl arrives as the others decline. The second panel contains no
trajectory that reaches that line at all; only Sampras spends appreciable time
above the looser one, between 1993 and 1997, and the three players who begin
the block at about that level, Lendl, Becker and Edberg, fall beneath it
within its first two years. The third is the one in which three trajectories
sit above the stricter threshold at the same time, from about 2011 until the
end of the block. In the fourth Djokovic stands far above it in 2015 and 2016
and then sits at about the line until 2023, while Federer and Nadal fall away
beneath it, and the young players who eventually replace them cross it only in
the closing years.

The composition of the top three differs more sharply than the levels do.
Between 1978 and 1989 six players hold those positions with any regularity:
John McEnroe in $74$ per cent of weeks, Ivan Lendl $70$, Jimmy Connors $56$,
Bj\"orn Borg $36$, Boris Becker $23$ and Stefan Edberg $21$. Between 2002 and
2013 three players account for almost all of it, Federer at $90$ per cent,
Nadal $74$ and Djokovic $51$, with Lleyton Hewitt next at $25$. Between 2014
and 2025 the concentration is greater still at the top and thinner below:
Djokovic holds a top-three position in $98$ per cent of weeks, Federer $62$ and
Nadal $52$, with Carlos Alcaraz and Jannik Sinner appearing only towards the
end at $29$ and $24$. The same three men occupy both later blocks, but not in
the same way, and Section~\ref{sec:concur} makes the difference precise.

\subsection{Peak relative strength}
 
Table~\ref{tab:prof} describes the top of the field one rank at a time. We turn
now to individual players, and in particular to the highest relative strength each
attained over his career, which is the quantity the popular debate concerns.
 
\begin{table}[t]
\centering
\small
\caption{Posterior peak relative strength over 1978--2025, in calibrated rating points
above the contemporaneous reference level. Entries are posterior medians with
$95\%$ credible intervals in parentheses, ordered by median reading down the
columns. \vcheck{script 04, Table 6 block}}
\label{tab:peaks}
\setlength{\tabcolsep}{3.5pt}
\begin{tabular}{lclclc}
\toprule
Player & peak & Player & peak & Player & peak \\
\midrule
Djokovic & 504 (448, 555) & Connors & 411 (366, 468) & Edberg & 323 (287, 370) \\
Borg & 461 (412, 523) & Sinner & 396 (327, 468) & Wilander & 308 (265, 358) \\
McEnroe & 453 (409, 511) & Murray & 368 (329, 412) & del Potro & 298 (258, 345) \\
Lendl & 449 (401, 507) & Alcaraz & 359 (298, 429) & Agassi & 294 (254, 336) \\
Federer & 440 (404, 493) & Becker & 344 (301, 399) & Hewitt & 264 (223, 308) \\
Nadal & 426 (384, 482) & Sampras & 333 (291, 381) & Roddick & 252 (218, 296) \\
\bottomrule
\end{tabular}
\end{table}
 
\begin{table}[t]
\centering
\small
\caption{Posterior probability that the row player's peak relative strength
exceeds that of the column player. \vcheck{script 04, Table 7 block}}
\label{tab:pairwise}
\begin{tabular}{lcccccc}
\toprule
& Djokovic & Federer & Nadal & Borg & McEnroe & Lendl \\
\midrule
Djokovic & --   & 0.94 & 0.98 & 0.85 & 0.88 & 0.88 \\
Federer  & 0.06 & --   & 0.69 & 0.27 & 0.36 & 0.37 \\
Nadal    & 0.02 & 0.31 & --   & 0.17 & 0.25 & 0.27 \\
Borg     & 0.15 & 0.73 & 0.83 & --   & 0.59 & 0.64 \\
McEnroe  & 0.12 & 0.64 & 0.75 & 0.41 & --   & 0.57 \\
Lendl    & 0.12 & 0.63 & 0.73 & 0.36 & 0.43 & --   \\
\bottomrule
\end{tabular}
\end{table}
 
\begin{figure}[t]
\centering
\includegraphics[width=0.5\textwidth]{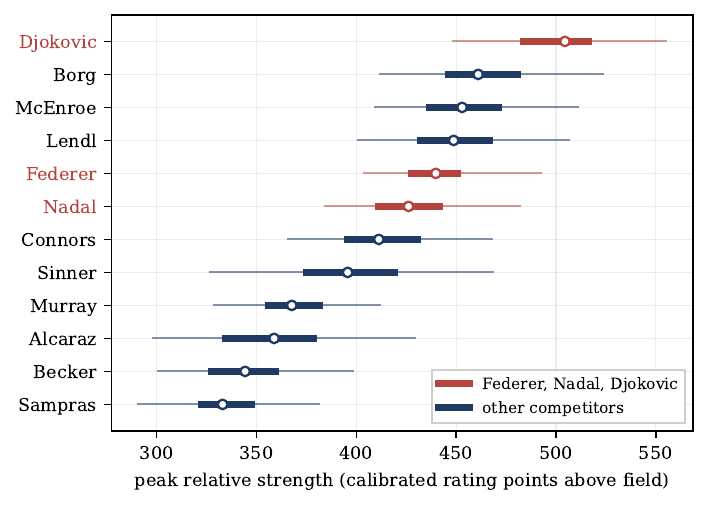}
\caption{Posterior distribution of peak relative strength. Thick bars denote
interquartile ranges and thin bars $95\%$ credible intervals. \vcheck{script 10, fig2}}
\label{fig:peaks}
\end{figure}

Table~\ref{tab:peaks}, Table~\ref{tab:pairwise} and Figure~\ref{fig:peaks}
report posterior peak relative strength. One player separates from the field. The posterior probability that Djokovic's peak exceeds that of any listed
rival lies between $0.85$ and $0.98$. No other separation is established. In
particular, the posterior probability that Federer's peak exceeds Borg's is $0.27$, and the
corresponding probability for Nadal is $0.17$, so that on this
criterion the leading players of the early 1980s are, if anything, ahead of two of the three players whose era is usually regarded as exceptional.
The credible intervals for Borg, McEnroe, Lendl, Federer and Nadal overlap
almost completely.
 
\subsection{Concurrent dominance}
\label{sec:concur}
 
\begin{table}[t]
\centering
\small
\caption{Concurrent dominance by era at two rate-calibrated thresholds. The
threshold is set within each posterior draw so that the mean exceedance count
over 1978--2025 equals $\bar N$. Total periods with $N_t\geq3$ are posterior
medians with $95\%$ credible intervals. \vcheck{script 04, Table 8 block}}
\label{tab:conc}
\setlength{\tabcolsep}{3.5pt}
\begin{tabular}{lrrrcr}
\toprule
Era & periods & $E[N_t]$ & mean $P(N_t\geq3)$ & periods $N_t\geq3$ & periods $p_t\geq0.5$ \\
\midrule
\multicolumn{6}{l}{\textit{$\bar N = 2$ players per period}}\\
1978--1989 & 537 & 2.79 & 0.639 & 341 (250, 434) & 315 \\
1990--2001 & 488 & 0.42 & 0.010 & 0 (0, 28) & 0 \\
2002--2013 & 455 & 2.21 & 0.418 & 191 (138, 233) & 194 \\
2014--2025 & 444 & 2.56 & 0.541 & 241 (163, 330) & 217 \\
\midrule
\multicolumn{6}{l}{\textit{$\bar N = 1$ player per period}}\\
1978--1989 & 537 & 1.56 & 0.106 & 51 (0, 137) & 0 \\
1990--2001 & 488 & 0.02 & 0.000 & 0 (0, 0) & 0 \\
2002--2013 & 455 & 1.23 & 0.136 & 63 (5, 115) & 57 \\
2014--2025 & 444 & 1.15 & 0.076 & 29 (0, 96) & 0 \\
\bottomrule
\end{tabular}
\end{table}
 
\begin{figure}[t]
\centering
\includegraphics[width=0.82\textwidth]{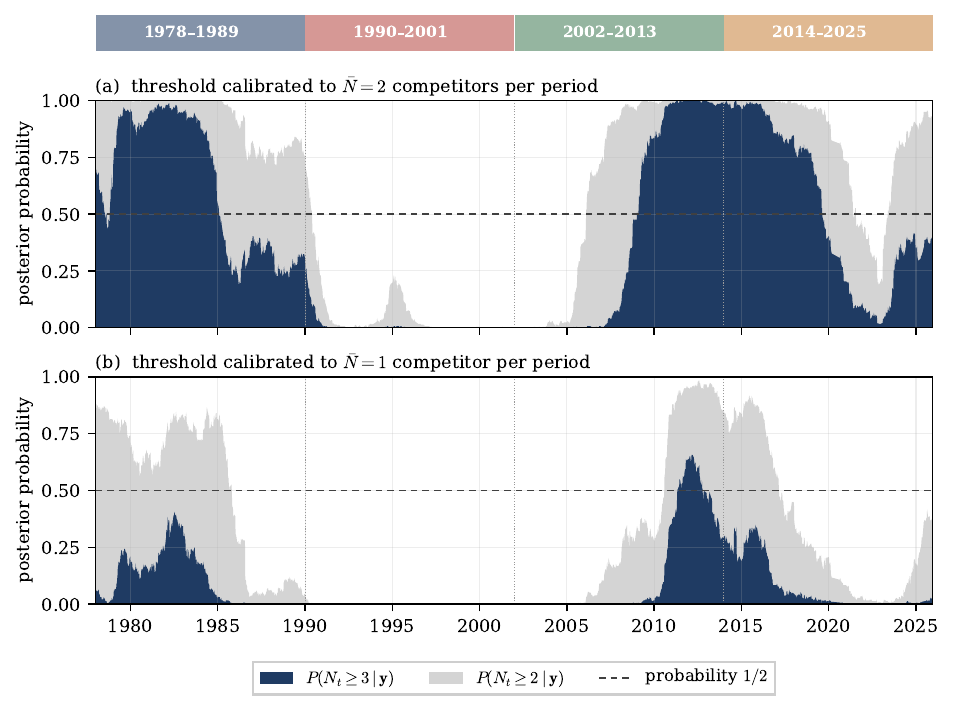}
\caption{Posterior probability that at least three players (dark) or at least
two (grey) are simultaneously dominant, through time, at the two
rate-calibrated thresholds. The dashed line marks probability one half; shaded
vertical bands denote the four blocks. \vcheck{script 10, fig1}}
\label{fig:conc}
\end{figure}

Peak strength describes players one at a time. The claim under examination,
however, concerns how many of them occupied the upper tail at once, which is
the exceedance count $N_t(u)$ of \eqref{eq:count}. Table~\ref{tab:conc} and
Figure~\ref{fig:conc} report it by era at the two rate-calibrated thresholds
of Section~\ref{sec:func}. Three findings emerge.
 
First, at the stricter threshold only one block contains any period in which
the simultaneous presence of three dominant players is more probable than not,
and it is 2002--2013, with $57$ such periods against none in any other block,
the 1980s included. At the looser threshold three blocks qualify, led by the
1980s with $315$ such periods against $217$ for 2014--2025 and $194$ for
2002--2013.
 
Second, the two blocks spanning the careers of the three players do not
resemble each other. At the looser threshold 2014--2025 is the stronger,
$0.541$ against $0.418$; at the stricter one the ordering reverses decisively,
$0.076$ against $0.136$. The first half had three players far clear of the
field, the second had a strong upper order but rarely three players that far
clear at once. The 1980s, meanwhile, lead at the looser threshold, $0.639$, and
trail 2002--2013 at the stricter, $0.106$ against $0.136$: more players
moderately clear of the field, fewer far clear of it. The advantage of
2002--2013 therefore appears only as the definition of dominance is made more
demanding, which is the pattern the persistence profiles sharpen below.
 
Third, the block from 1990 to 2001 is a structural anomaly. Its expected
exceedance count is $0.42$ at the looser threshold and $0.02$ at the stricter
one, against values between $1.1$ and $2.8$ in every other block. On every
measure at every threshold in every specification we examined, the upper tail
of the relative strength distribution was unoccupied for twelve years.
 
\subsection{Persistence of concurrent dominance}
\label{sec:persist}
 
We describe persistence through the lag profile $\{\pi_k\}$ and the block
profile $\{\beta_w\}$ rather than through a run length, for the reasons given in
Section~\ref{sec:func}.
 
\begin{table}[t]
\centering
\small
\caption{Persistence profiles by era at the stricter rate-calibrated threshold
$\bar N=1$. Entries are posterior medians. The 1990--2001 block admits too few
exceedances for the conditional profile to be defined. \vcheck{script 04, pi\_k and beta\_w rows at mean N=1}}
\label{tab:persist}
\setlength{\tabcolsep}{4pt}
\scalebox{0.82}{%
\begin{tabular}{lcccccccccc}
\toprule
& \multicolumn{5}{c}{lag persistence $\pi_k$} & \multicolumn{5}{c}{block persistence $\beta_w$} \\
\cmidrule(lr){2-6}\cmidrule(lr){7-11}
Era & $k=1$ & $k=4$ & $k=13$ & $k=26$ & $k=52$ & $w=4$ & $w=13$ & $w=26$ & $w=52$ & $w=104$ \\
\midrule
1978--1989 & 0.89 & 0.72 & 0.48 & 0.28 & 0.12 & 0.07 & 0.03 & 0.00 & 0.00 & 0.00 \\
1990--2001 & --   & --   & --   & --   & --   & 0.00 & 0.00 & 0.00 & 0.00 & 0.00 \\
2002--2013 & 0.93 & 0.80 & 0.67 & 0.54 & 0.44 & 0.11 & 0.06 & 0.02 & 0.00 & 0.00 \\
2014--2025 & 0.87 & 0.66 & 0.37 & 0.16 & 0.06 & 0.04 & 0.01 & 0.00 & 0.00 & 0.00 \\
\bottomrule
\end{tabular}}
\end{table}
 
\begin{figure}[t]
\centering
\includegraphics[width=0.82\textwidth]{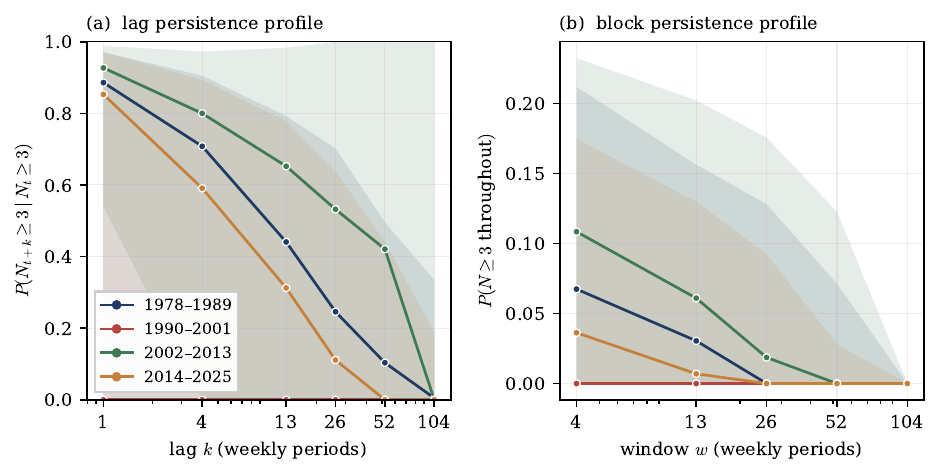}
\caption{Lag and block persistence profiles by era at the stricter
rate-calibrated threshold, with $95\%$ credible bands. Horizontal axes are
logarithmic. \vcheck{script 11, fig7}}
\label{fig:persist}
\end{figure}

Table~\ref{tab:persist} and Figure~\ref{fig:persist} report the profiles, and
they separate the blocks far more sharply than the rates do. At a lag of one
week all three populated blocks agree to within six points, between $0.87$ and
$0.93$. At a lag of one year they do not: the probability that three-way
dominance survives is $0.44$ for 2002--2013, $0.12$ for 1978--1989 and $0.06$
for 2014--2025. The block profile tells the same story, with $\beta_{26}=0.02$
for 2002--2013 against zero elsewhere. The rate functionals of
Table~\ref{tab:conc} do not contain this, their spread across the same three
blocks running only from $0.076$ to $0.136$: the blocks differ much more in how
long dominance was sustained than in how often it occurred, and on that
criterion 2002--2013 stands alone, with 2014--2025 below even the 1980s.
 
\subsection{Duration of individual dominance}
 
\begin{table}[t]
\centering
\small
\caption{Posterior sojourn above the dominance threshold at $\bar N=2$, in
weekly periods, with $95\%$ credible intervals, ordered reading down the
columns. \vcheck{script 04, Table 10 block}}
\label{tab:sojourn}
\setlength{\tabcolsep}{3.5pt}
\begin{tabular}{lclclc}
\toprule
Player & sojourn & Player & sojourn & Player & sojourn \\
\midrule
Federer & 612 (526, 714) & Lendl & 408 (358, 448) & Murray & 215 (118, 302) \\
Djokovic & 578 (457, 657) & McEnroe & 365 (298, 439) & Borg & 205 (148, 295) \\
Nadal & 510 (396, 598) & Connors & 305 (215, 369) & Becker & 116 (25, 206) \\
& & & & Sampras & 113 (14, 209) \\
\bottomrule
\end{tabular}
\end{table}
 
The profiles above describe the field as a whole. The corresponding quantity
for an individual player is his sojourn, the total number of periods in which
his relative strength exceeded the threshold, which is a rate functional and
therefore among the reliably estimable summaries. Table~\ref{tab:sojourn}
reports it. The three players of the later era occupy the first three
positions; the intervals of the first two exclude the fourth-placed player,
while Nadal's overlaps Lendl's. Table~\ref{tab:peaks} orders the same players
differently, so the two tables together locate where the later era differs:
not in peak strength, but in how long the upper tail was occupied.
 
Borg records the second-highest peak in the sample and the eighth-longest
sojourn, having retired at twenty-six. His interval overlaps Murray's, so the
ordering between those two should not be read as settled.
 
\subsection{The strongest decade against the strongest other decade}
\label{sec:best}
 
The comparison that requires no division into blocks searches over every
ten-year window in the record for the one with the highest mean
$P(N_t\geq3)$, then for the best window disjoint from it, and compares the
two, so that each period is judged on its strongest decade.
 
\begin{table}[t]
\centering
\small
\caption{The ten-year window with the highest mean $P(N_t\geq3)$, and the best
window disjoint from it, at each rate-calibrated threshold.
\vcheck{script 17}}
\label{tab:best}
\resizebox{\textwidth}{!}{%
\begin{tabular}{llccccc}
\toprule
& Window & $E[N_t]$ & mean $P(N_t\geq3)$ & periods $N_t\geq3$ & periods $p_t\geq0.5$ & $\pi_{52}$ \\
\midrule
\multicolumn{7}{l}{\textit{$\bar N = 1$}}\\
strongest & 2010--2019 & 1.86 & 0.246 & 93 & 57 & 0.29 \\
best other & 1978--1987 & 1.73 & 0.125 & 51 & 0 & 0.12 \\
\midrule
\multicolumn{7}{l}{\textit{$\bar N = 2$}}\\
strongest & 2009--2018 & 3.48 & 0.911 & 356 & 385 & 0.96 \\
best other & 1978--1987 & 2.93 & 0.708 & 324 & 315 & 0.80 \\
\bottomrule
\end{tabular}}
\end{table}
 
Table~\ref{tab:best} reports the outcome, and it is the same at both
thresholds. The strongest decade in the record falls in the late 2000s and
2010s, and the strongest decade disjoint from it is the opening of the window,
1978--1987. At the demanding threshold the later decade has twice the mean
concurrence, $0.246$ against $0.125$, and $57$ periods in which three-way
dominance is more probable than not against none. At the permissive threshold
it has $0.911$ against $0.708$.
 
The difference is one of kind rather than degree. In 2010--2019 three players
held a top-three position in $99$, $92$ and $80$ per cent of weeks, Djokovic,
Federer and Nadal, with Murray next at $26$. In 1978--1987 four players share
the same positions less completely, McEnroe at $84$ per cent, Connors $67$,
Lendl $65$ and Borg $43$. The earlier decade had four strong players trading
places; the later had three occupying them at once.
 
\subsection{Dominance across playing surfaces}
\label{sec:surf}

All results so far pool matches across playing surfaces. We now ask whether
occupancy of the upper tail transfers between them, using the coefficients of
\eqref{eq:chieta}. This requires the surface-specific specialization of
\eqref{eq:state}, with $S=3$, which we fit afresh and which answers the
question in a single fit rather than through three independent per-surface
models. Restricting to the $143{,}383$ matches played on the three principal
surfaces, involving $4{,}970$ players over $1{,}818$ periods, the
cross-surface innovation correlation is estimated at $\rho=0.90$ at an
interior maximum, with $\tau=0.05$ and predictive accuracy $0.667$. The
improvement in prequential log-likelihood over the independent specification
$\rho=0$ is $1{,}365$ units, so the correlation is sharply identified. The
value implies that about a tenth of the variance of a week's change in
strength is specific to the surface on which the results were earned, and
since the criterion is predictive, a correlation that did not improve the
prediction of matches before they were played would not have been selected,
so the approximation of Section~\ref{sec:approx} cannot manufacture it.
Strength development is therefore strongly but not completely shared across surfaces, so neither a single common strength nor three separate analyses would be adequate. We summarize career peak strengths through their ranks within the population
of $199$ players with at least forty matches on each surface, and compute the
dependence coefficients of \citet{ledford1996} within each posterior draw.
 
\begin{table}[t]
\centering
\small
\caption{Cross-surface upper tail dependence under the joint model, posterior
medians with $95\%$ credible intervals. \vcheck{script 05, Table 12 block}}
\label{tab:surf}
\resizebox{\textwidth}{!}{%
\begin{tabular}{lcccc}
\toprule
& \multicolumn{2}{c}{$q=0.90$} & \multicolumn{2}{c}{$q=0.95$} \\
\cmidrule(lr){2-3}\cmidrule(lr){4-5}
Pair & $\chi$ & $\eta$ & $\chi$ & $\eta$ \\
\midrule
Hard--clay  & 0.60 (0.50, 0.70) & 0.82 (0.77, 0.87) & 0.60 (0.40, 0.70) & 0.86 (0.77, 0.89) \\
Hard--grass & 0.65 (0.50, 0.75) & 0.84 (0.77, 0.89) & 0.60 (0.40, 0.80) & 0.86 (0.77, 0.93) \\
Clay--grass & 0.50 (0.35, 0.60) & 0.77 (0.69, 0.82) & 0.40 (0.30, 0.60) & 0.77 (0.71, 0.86) \\
\midrule
Extreme on all three & \multicolumn{2}{c}{8 (6, 11)} & \multicolumn{2}{c}{4 (3, 5)} \\
Under independence   & \multicolumn{2}{c}{0.20} & \multicolumn{2}{c}{0.02} \\
Under perfect dependence & \multicolumn{2}{c}{20} & \multicolumn{2}{c}{10} \\
\bottomrule
\end{tabular}}
 
\end{table}
 
\begin{table}[t]
\centering
\small
\caption{Posterior probability of being extreme on all three surfaces
simultaneously, and posterior median surface rank percentiles, under the joint
model. \vcheck{script 05, Table 13 block; confirm del Potro and Wilander rows}}
\label{tab:surfmemb}
\begin{tabular}{lcccccc}
\toprule
& \multicolumn{2}{c}{$P(\text{extreme on all three})$} & \multicolumn{3}{c}{median rank percentile} \\
\cmidrule(lr){2-3}\cmidrule(lr){4-6}
Player & $q=0.90$ & $q=0.95$ & hard & clay & grass \\
\midrule
Djokovic    & 1.00 & 1.00 & 0.995 & 0.990 & 0.995 \\
Federer     & 1.00 & 1.00 & 0.990 & 0.985 & 0.990 \\
Nadal       & 1.00 & 0.97 & 0.978 & 0.995 & 0.980 \\
Murray      & 1.00 & 0.91 & 0.980 & 0.970 & 0.985 \\
McEnroe     & 0.70 & 0.12 & 0.960 & 0.925 & 0.968 \\
del Potro   & 0.67 & 0.12 & 0.955 & 0.950 & 0.930 \\
Lendl       & 0.57 & 0.12 & 0.980 & 0.975 & 0.915 \\
Hewitt      & 0.32 & 0.01 & 0.930 & 0.885 & 0.960 \\
Soderling   & 0.24 & 0.01 & 0.905 & 0.940 & 0.895 \\
Sampras     & 0.23 & 0.01 & 0.970 & 0.870 & 0.965 \\
\bottomrule
\end{tabular}
\end{table}
 
\begin{figure}[t]
\centering
\includegraphics[width=0.82\textwidth]{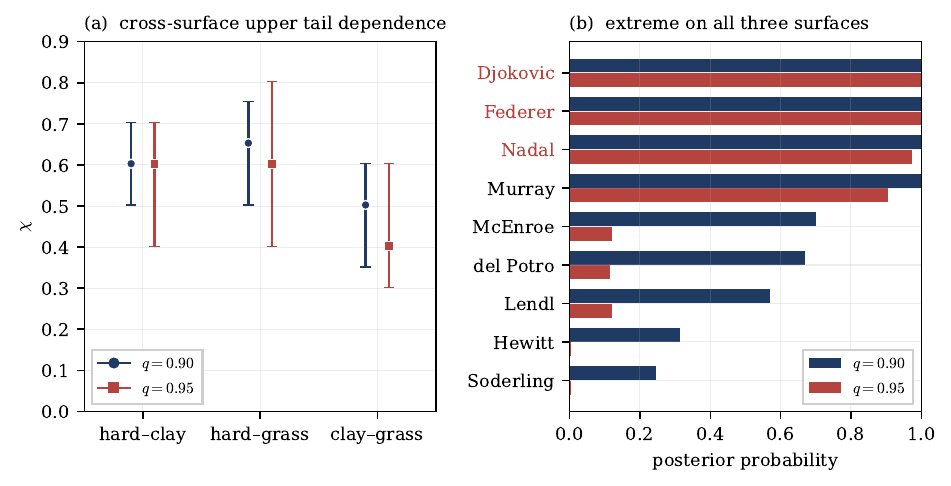}
\caption{Left: posterior cross-surface upper tail dependence coefficients under
the joint model, with credible intervals. Right: posterior probability of
occupying the upper tail on all three surfaces simultaneously. \vcheck{script 10, fig5; reads out/05\_surfaces.pkl}}
\label{fig:surf}
\end{figure}

Tables~\ref{tab:surf} and \ref{tab:surfmemb} and Figure~\ref{fig:surf} report
the results. Upper tail dependence is substantial but incomplete, and the weakest link is between clay and grass, where $\chi$ falls to $0.40$ at the more
demanding level. The number of players extreme on all three surfaces is
eight at $q=0.90$ and four at $q=0.95$, against expectations of $0.20$ and
$0.02$ under independence and $20$ and $10$ under perfect dependence. At the more demanding level, four players have posterior probability above one
half of occupying the upper tail on every surface: Djokovic and Federer at
unity, Nadal at $0.97$ and Murray at $0.91$. Surface universality is therefore a
rare property, shared by all three of the players under study together with one
contemporary, and by no predecessor with probability above $0.12$.
 
Fitting three separate models to the same data understates this conclusion. Under separate fits Nadal's posterior probability at $q=0.95$ was $0.32$ rather than
$0.97$, and his grass percentile $0.940$ rather than $0.980$, because grass
matches are the scarcest of the three and an independent fit cannot borrow the
information his hard and clay records supply. With $\rho$ estimated at $0.90$ the borrowing is substantial and the estimates are correspondingly sharper.

\section{Discussion}
\label{sec:disc}

\subsection{Comparison with previous findings}

Three earlier answers to the question of the strongest player of the open
era can be set against Table~\ref{tab:peaks}. \citet{bakermchale2014} found
Federer the most likely candidate, with Borg and Connors close behind, and
noted that their rankings largely agreed with the count of major titles.
Their record closes in the early 2010s, before most of Djokovic's career, so
the one separation we establish could not have appeared in it; Borg, on our
evidence, is if anything ahead of Federer rather than behind him, with
posterior probability $0.73$. The agreement with title counts is where we
part company. Borg, McEnroe and Lendl won $26$ majors between them against
$66$ for Federer, Nadal and Djokovic, yet on peak the five players other than
Djokovic cannot be told apart. A title count accumulates over a career, so it
is closer to a sojourn than to a peak, and on sojourn our ordering does agree
with it. The same distinction accounts for \citet{radicchi2011}, whose
cumulative prestige score placed Connors first, Federer seventh and Borg
tenth: Borg is second on peak and eighth on sojourn, having retired at
twenty-six, and Federer fifth and first. The lists disagree because the
quantities do. \citet{bakermchale2017} found for the women's tour that
intervals on the rankings left several players in contention, and
Table~\ref{tab:pairwise} says the same of the men.

The closest antecedent on eras is \citet{breznik2025}, who identify five
strong eras over a little more than thirty years of data, the most recent
ending in July 2019, with Federer the player present in most of them. On both
points we agree: the strongest decade in Table~\ref{tab:best} ends in 2018 or
2019 depending on the threshold, and Federer's sojourn is the longest in
Table~\ref{tab:sojourn}. Their record does not reach the early 1980s,
however, which is where our evidence places the precedent, at least in how
many players were dominant at once. An era in their analysis is, moreover,
either present or absent, whereas the distinction that matters most here,
between 2002--2013 and 2014--2025 in the chance that three-way dominance
survives a year, is one of degree.

\citet{bayrambarla2026} report from the same match record that the sport has
become more competitive, with more frequent upsets and top-player dominance
spread more evenly across the field. This is not in tension with our findings
once threshold dependence is taken into account: the body of the distribution
can compress while its extreme tail concentrates, and at the permissive
threshold the 1980s lead every later block while at the demanding one
2002--2013 does. Table~\ref{tab:prof} adds that the change was not monotone,
since the top of the field stood lower above the rest in 1990--2001 than in
any block before or since. \citet{radicchi2011} named Sampras the best player
of 1991--2000, and his peak of $333$ points, more than a hundred below the
leaders of the blocks on either side, is that compression seen from the top.

Finally, our comparisons are of gaps, not levels. \citet{berry1999} bridge
eras in order to compare all players directly, which requires assumptions
about how the talent pool changes; the relative strength of \eqref{eq:rel}
measures each player against the tenth to hundredth ranked players of his own
day. That the top of
the field stood higher above the rest in the 1980s than in the 2000s is
therefore compatible with the later field being stronger throughout, as the
increase in competitiveness just noted suggests. Whether Borg at his peak
would have beaten Djokovic at his is not a question the match record settles,
and we have not asked it.

\subsection{Extremes of inferred strengths}

Extreme value methods in sport have been applied to measured performances:
ultimate records \citep{einmahlmagnus2008,einmahlhe2026},
rankings of performances \citep{stephensontawn2013} and swim times
\citep{spearing2021}. Tennis produces no such measurement. A match yields a
winner, and relative strength is a latent variable of a statistical model; we
are not aware of a previous application of the peaks-over-threshold apparatus
to a sport of that kind. Three things change: the threshold is calibrated to
a rate, so that eras are compared at equal rarity; exceedances are dependent
across players by construction, and the count $N_t(u)$ absorbs that dependence
into a single process, the estimand being sustained joint occupation of the
tail rather than the concurrence of maxima, as Section~\ref{sec:lit} noted; and
the
dependence within a player that \citet{spearingtawn2023} model directly for
measured performances is here induced by the state equation \eqref{eq:state},
so that its standard tail summary, the extremal index, cannot be estimated
from inferred strengths.

The last point is easily misread, since estimation aggravates the difficulty
without causing it. The longest run is not qualitatively robust
\citep{hampel1971}: it varies by half its own size across replicates of one
process even when strengths are known exactly, and estimation then supplies
the interruptions to which it is sensitive. Posterior sampling reduces the
inflation only from $1.75$ to $1.32$, so the remedy is a change of summary
rather than of estimator. Once it is made, Section~S5 of the supplementary
material shows that on the
tennis data the value obtained by treating ratings as observed falls inside
the posterior interval at both thresholds.
 
\subsection{Limitations}
 
Four limitations deserve statement. First, sojourn above a dominance
threshold conflates playing strength with career duration, and career duration
depends on factors, including advances in the medical management of athletes,
that lie outside the model. The comparison between Borg's peak and his sojourn
is a direct illustration; the limitation is one of the estimand rather than
of estimation, and it should temper the reading of Table~\ref{tab:sojourn}. Second, our analysis concerns the men's professional tour. The motivating
question is specific to that population, and we make no claim about whether
the same structure obtains elsewhere. Third, the reference level in \eqref{eq:rel} is one of several defensible
choices. Section~S4 of the supplementary material reports the effect of
widening and narrowing the
excluded group: peak strengths shift by up to $20$ points, but no ordering and
no era comparison is disturbed. Fourth, the cross-surface correlation is held
constant over a period in which the surfaces themselves have converged in
pace \citep{bayrambarla2026}; a correlation allowed to drift would probably
rise over the record, and the results of Section~\ref{sec:surf} should be read
with that in mind.

\section{Conclusions}
\label{sec:conc}

Applied to the complete record of men's professional tennis, the analysis
supports the conventional account of the recent era in one respect and not
in others. Only Djokovic separates from the field on peak strength, Federer
and Nadal being indistinguishable from Borg, McEnroe and Lendl; what
distinguishes the period is how long three players occupied the upper tail
together, and even that holds for 2002--2013 rather than throughout, the
probability that three-way dominance survives a year being $0.44$ then
against $0.12$ in the early 1980s and $0.06$ between 2014 and 2025. The
sharper anomaly is the twelve years from 1990, in which the upper tail was
not merely unoccupied but compressed. Each rests on summaries that average
over fixed windows rather than on the run lengths that usually carry such
claims, and Section~\ref{sec:sim} is our reason for insisting on the
difference. The structure is not peculiar to sport. A volatility filtered from
asset returns or the condition of a machine inferred from sensor readings is
a latent process that is thresholded and then described by how long it stayed
beyond the threshold, and the same choice between run-based and fixed-window
summaries arises in each. Whether the distinction holds as generally as that
is the question we would take up next.

\section*{Data and code}

All results are computed from publicly available match records. Data and code
to replicate every table and figure are available at
\url{https://github.com/manueleleonelli/Tennis_Extremes}. 

\section*{Acknowledgements}

This work was supported by the Spanish Agencia Estatal de Investigaci\'on
grant PID2023-153222OB-I00.

\section*{Declaration of generative AI and AI-assisted technologies in the
manuscript preparation process}

During the preparation of this work the author used generative AI tools to
support coding and writing. After using these tools, the author reviewed and
edited the content as needed and takes full responsibility for the content of
the published article.

\bibliographystyle{elsarticle-harv}
\bibliography{refs}

\newpage 

\appendix

\section{Fitted hyperparameters and calibration}
\label{app:hyper}
 
\begin{table}[h]
\centering
\caption{Selected hyperparameters of the model of Section~4 of the main paper.
\vcheck{script 02}}
\label{tab:hyper}
\begin{tabular}{lr}
\toprule
Innovation standard deviation $\tau$ & 0.0473 \\
Posterior standard deviation of $\tau$ & 0.00080 \\
Initial standard deviation $\sigma_0$ & 1.0 \\
Discrimination for best-of-five matches $a_5$ & 1.15 \\
Cross-surface innovation correlation $\rho$ & 0.90 \\
Recalibration slope & 0.851 \\
Posterior draws & 300 \\
\bottomrule
\end{tabular}
\end{table}
 
Table~\ref{tab:hyper} collects the values selected by the procedure of
Section~4.3 of the main paper. Fitting a quadratic to a fine grid in $\tau$ gives a
mode of $0.0473$ with an implied posterior standard deviation of $0.00080$ and
a $95\%$ interval of $[0.0457, 0.0489]$, so that any weakly informative prior
would be numerically irrelevant at this sample size. The discrimination for
best-of-five matches is estimated at $1.15$, confirming that the longer match
is more discriminating, though the improvement in fit is modest.
 
\begin{figure}[h]
\centering
\includegraphics[width=0.48\textwidth]{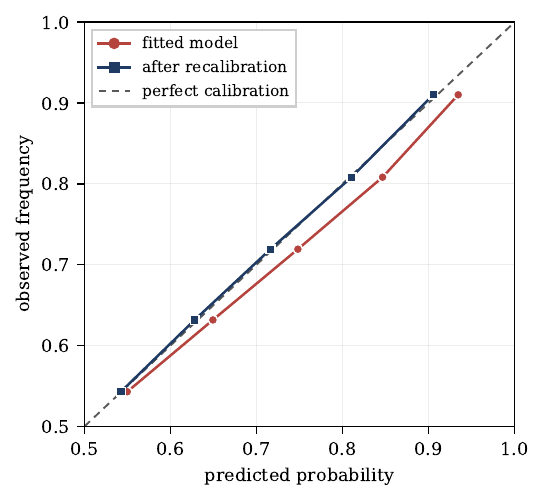}
\caption{Reliability of the fitted model before and after recalibration on the
logit scale. The estimated correction is a pure rescaling, with intercept zero
and slope $0.851$. \vcheck{script 10, fig4; reads out/02\_hyper.json}}
\label{fig:rel}
\end{figure}

The overconfidence noted in Section~5.2 of the main paper might be addressed
by adding a dispersion parameter to the link, equation~(5) there, and it is worth recording why
this does not work. Such a parameter is identified only jointly with the scale
of the latent process, so the optimization runs along a ridge, gaining $0.002$
log-likelihood units per match while migrating without settling. Recalibration
after fitting diagnoses the same quantity cleanly. Figure~\ref{fig:rel} shows
the reliability diagram before and after.
 
\section{The within-period update}
\label{app:update}

Section~4.4 of the main paper approximates the posterior by a mean-field step
across players and a Laplace step within each player. This section gives the
resulting update equations. Because each match involves a single component of the state, the within-period
problem is univariate. Write $m_{i,t}$ and $P_{i,t}$ for the prior mean and
variance of the relevant component, carried into period $t$ after the
propagation implied by equation~(4) of the main paper. Let $y_{ij}$ equal one if $i$ beat $j$
and zero otherwise, and let
\begin{equation*}
g_j = a_f\big(1+3P_{j,t}/\pi^2\big)^{-1/2}
\end{equation*}
be the attenuation induced by uncertainty about the opponent's strength, the
device introduced by \citet{glickman1999} for propagating an approximately
Gaussian belief through a logistic link. The mode $\hat\theta_{i,t}$ of the period posterior solves
\begin{equation}
S_{i,t}(\theta) = \frac{\theta-m_{i,t}}{P_{i,t}},
\label{eq:map}
\end{equation}
where $S_{i,t}(\theta)=\sum_{j} g_j\{y_{ij}-E_{ij}(\theta)\}$ is the score
contributed by the matches $i$ played in period $t$, and
$E_{ij}(\theta)=[1+\exp\{-g_j(\theta-\hat\theta_{j,t})\}]^{-1}$ is the implied
win probability against opponent $j$.

Equation~\eqref{eq:map} has no closed-form solution. The Glicko rating update
takes a single Fisher scoring step from the prior mean, whose accuracy degrades
when $S_{i,t}$ is large; as noted in Section~4.4 of the main paper, we solve
\eqref{eq:map} by Newton's method instead. Writing $I_{i,t}=\sum_j g_j^2 E_{ij}(1-E_{ij})$ for the
observed information at the converged mode, the period posterior variance is
$\hat P_{i,t}=(P_{i,t}^{-1}+I_{i,t})^{-1}$.

To recover a linear Gaussian representation suitable for smoothing, note that
the Laplace approximation to the period posterior is reproduced exactly by a
Gaussian pseudo-observation $y_{i,t}$ with variance $v_{i,t}=I_{i,t}^{-1}$,
where
\begin{equation*}
y_{i,t} = \hat\theta_{i,t} + \frac{v_{i,t}}{P_{i,t}}\big(\hat\theta_{i,t}-m_{i,t}\big).
\end{equation*}
Writing
$\mathbf{K}=\mathbf{P}_{i,t}\mathbf{e}_s\,(\mathbf{e}_s^{\top}\mathbf{P}_{i,t}\mathbf{e}_s+v_{i,t})^{-1}$
for the Kalman gain, where $\mathbf{e}_s$ selects the surface on which the match
was played, the full state is updated by
\begin{equation*}
\mathbf{m}_{i,t}\leftarrow \mathbf{m}_{i,t}+\mathbf{K}\big(y_{i,t}-\mathbf{e}_s^{\top}\mathbf{m}_{i,t}\big),
\qquad
\mathbf{P}_{i,t}\leftarrow \mathbf{P}_{i,t}-\mathbf{K}\,\mathbf{e}_s^{\top}\mathbf{P}_{i,t},
\end{equation*}
so that a result on one surface revises the estimated strength on the others in
proportion to $\rho$. Under the scalar specialization this step is vacuous and
the recursion reduces to the univariate update above.

\section{Accuracy of the posterior approximation}
\label{app:valid}
 
The approximations of Section~4 of the main paper replace an intractable posterior
with a tractable one, and their adequacy is an empirical question. We answer it
by constructing an exact sampler for the posterior, equation~(6) of the main
paper, and comparing the two
on a subsample small enough for the exact scheme to be run to convergence.
 
The exact scheme uses the P\'olya-Gamma data augmentation of
\citet{polson2013}, written PG below. Introducing latent variables
$\omega_m\sim\mathrm{PG}(1,\psi_m)$ with
$\psi_m=\theta_{i_m,t_m}-\theta_{j_m,t_m}$ renders the logistic likelihood
conditionally Gaussian in the strengths, since
\begin{equation*}
\frac{\exp(\kappa_m\psi_m)}{1+\exp(\psi_m)}
\;\propto\;
E_{\omega_m}\exp\Big\{\kappa_m\psi_m-\tfrac{1}{2}\omega_m\psi_m^2\Big\},
\qquad \kappa_m=y_m-\tfrac12 .
\end{equation*}
Conditionally on $\{\omega_m\}$ and on the trajectories of all other players,
each player's trajectory follows a linear Gaussian state space model with
pseudo-observations $\theta_{j_m,t_m}+\kappa_m/\omega_m$ of precision
$\omega_m$, and can be drawn exactly by the same backward-sampling recursion
used above. Cycling over players in random order and alternating with the
P\'olya-Gamma draw yields a Gibbs sampler targeting (6) without
approximation. The scheme makes no mean-field simplification: the coupling across players induced by (6) is respected exactly, at the cost of requiring many sweeps.
 
We apply this to a subsample of the tennis records, described in full in
Section~5.1 of the main paper, comprising all matches between 2000 and 2010 among
players with at least $120$ matches in that window. This gives $21{,}163$
matches among $177$ players over $422$ weekly periods. We run $1{,}200$
iterations and discard the first $400$.
 
\begin{table}[t]
\centering
\caption{The approximate posterior compared against the exact P\'olya-Gamma
sampler. Quantities are computed on the relative scale of equation~(1) of the
main paper, which
is the scale on which all results are reported. \vcheck{script 07, the centred block}}
\label{tab:valid}
\resizebox{\textwidth}{!}{%
\begin{tabular}{lc}
\toprule
Correlation of posterior means across player-periods & 0.997 \\
Mean absolute difference in posterior mean & 7.2 points \\
Ninetieth percentile of absolute difference & 15.2 points \\
Ratio of posterior standard deviations, approximate to exact & 1.000 \\
\midrule
Correlation of posterior medians of peak strength & 0.998 \\
Mean absolute difference in posterior median of peak strength & 6.6 points \\
Ratio of posterior standard deviations of peak strength & 1.007 \\
\bottomrule
\end{tabular}}
 
\end{table}
 
Table~\ref{tab:valid} reports the comparison. Agreement is close. Posterior means
correlate at $0.997$ across player-periods and differ by $7.2$ rating points on
average, against posterior standard deviations of roughly $45$ points, and the
approximate posterior reproduces the exact posterior spread to three decimal
places.
 
One qualification applies. On the raw scale of the latent strengths
the approximate posterior means are displaced downward relative to the exact ones
by roughly $40$ points, an artefact of the mean-field treatment of opponents, and
the approximate posterior standard deviations are about six per cent too small.
Both effects are almost entirely common across players and therefore cancel under
the centring in (1). This is a further reason, beyond the drift in the
reference level discussed in Section~5.1 of the main paper, to conduct the
analysis on
relative rather than absolute strengths.
 
\section{Sensitivity analyses}
\label{app:sens}
 
The reference level in equation~(1) of the main paper excludes the nine
strongest active players
so that the very players under study do not set the standard against which they
are judged. The choice is a compromise: excluding too few contaminates the
reference, while excluding too many measures dominance against a field that
excludes everyone who might plausibly be called dominant. We therefore repeat
the analysis with the reference band running from the fifth to the hundredth
ranked player, which excludes only four, and from the fifteenth to the
hundredth, which excludes fourteen.
 
\begin{table}[t]
\centering
\caption{Sensitivity to the reference band used in equation~(1) of the main
paper. Peak relative
strengths are posterior medians in calibrated rating points; the remaining rows
are posterior medians of the quantities named. \vcheck{script 08, Table 14 block}}
\label{tab:sens}
\scalebox{0.85}{
\begin{tabular}{lccc}
\toprule
Reference band & 5--100 & 10--100 & 15--100 \\
\midrule
\multicolumn{4}{l}{\textit{Peak relative strength}}\\
Djokovic & 493 & 504 & 513 \\
Borg     & 451 & 461 & 468 \\
McEnroe  & 444 & 453 & 460 \\
Lendl    & 439 & 449 & 456 \\
Federer  & 432 & 440 & 447 \\
Nadal    & 415 & 426 & 434 \\
Sampras  & 325 & 333 & 340 \\
\midrule
\multicolumn{4}{l}{\textit{Posterior probability of ordering}}\\
$P(\text{Djokovic}>\text{Federer})$ & 0.92 & 0.94 & 0.94 \\
$P(\text{Djokovic}>\text{Borg})$    & 0.83 & 0.85 & 0.84 \\
$P(\text{Federer}>\text{Borg})$     & 0.28 & 0.27 & 0.26 \\
$P(\text{Nadal}>\text{Borg})$       & 0.17 & 0.17 & 0.17 \\
\midrule
\multicolumn{4}{l}{\textit{Mean $P(N_t\geq3)$, threshold $\bar N=1$}}\\
1978--1989 & 0.105 & 0.106 & 0.106 \\
1990--2001 & 0.000 & 0.000 & 0.000 \\
2002--2013 & 0.128 & 0.136 & 0.139 \\
2014--2025 & 0.072 & 0.076 & 0.082 \\
\midrule
\multicolumn{4}{l}{\textit{Mean $P(N_t\geq3)$, threshold $\bar N=2$}}\\
1978--1989 & 0.639 & 0.639 & 0.640 \\
1990--2001 & 0.011 & 0.010 & 0.009 \\
2002--2013 & 0.417 & 0.418 & 0.418 \\
2014--2025 & 0.545 & 0.541 & 0.538 \\
\midrule
\multicolumn{4}{l}{\textit{Lag persistence $\pi_{52}$, threshold $\bar N=1$}}\\
1978--1989 & 0.11 & 0.12 & 0.13 \\
1990--2001 & --   & --   & --   \\
2002--2013 & 0.41 & 0.44 & 0.47 \\
2014--2025 & 0.03 & 0.06 & 0.07 \\
\bottomrule
\end{tabular}}
\end{table}
 
Table~\ref{tab:sens} reports the comparison. Widening the excluded group raises
every peak relative strength, as it must, since a reference computed from weaker
players sits lower. The shift is close to uniform, between $7$ and $11$ points
per step, so the orderings are unaffected: the posterior probability
that Djokovic's peak exceeds Federer's moves from $0.92$ to $0.94$, and the
probability that Federer's exceeds Borg's from $0.28$ to $0.26$.
 
The concurrence and persistence results are equally stable, and at the looser
threshold essentially invariant: the mean probability of three-way dominance
moves from $0.639$ to $0.640$ in 1978--1989, from $0.417$ to $0.418$ in
2002--2013 and from $0.545$ to $0.538$ in 2014--2025. At the stricter
threshold 1978--1989 gives $0.105$ to $0.106$, 2002--2013 $0.128$ to $0.139$
and 2014--2025 $0.072$ to $0.082$. The one-year lag persistence moves from
$0.41$ to $0.47$ in 2002--2013 and from $0.11$ to $0.13$ in 1978--1989, so the
ratio between them, which carries the substantive claim, lies between $3.6$
and $3.7$ under every band. Excluding fewer players narrows the gap slightly,
as expected, since the fifth to ninth ranked players of a strong era are
themselves comparatively strong and raise the reference. The effect is small
enough not to disturb any conclusion.
 
A second choice worth testing is the format effect of equation~(5) of the main
paper. Because
best-of-five matches are more discriminating, a given win rate implies a
smaller strength difference when achieved over five sets than over three, so an
era containing proportionally more of them would have its strengths compressed.
The mix does drift, and it drifts against the later era rather than in its
favour. Refitting with
$a_5=1$, which removes the effect entirely, raises every relative strength by
roughly three per cent, but does so almost uniformly across periods: averaged
over the top five positions the increase is close to $3.2$ per cent in every
block, the spread across blocks being about a tenth of a percentage point. The
differential between eras is thus around a tenth of a percentage point, against
gaps of several per cent between the quantities being compared. The format
effect therefore acts as a global rescaling rather than as an advantage to any
period. This is not because the mix is constant, since it is not, but because
the effect is small: fifteen per cent additional discrimination applied to
roughly a fifth of matches leaves relative strengths moving by well under one
per cent even when the mix shifts by several percentage points. The likelihood
prefers $a_5=1.15$, but only slightly, the predictive accuracy falling from
$0.6745$ to $0.6742$ when the effect is removed.
 
\section{Comparison against treating ratings as observed}
\label{app:plugin}
 
\begin{table}[t]
\centering
\caption{Ratings treated as observed, against full posterior propagation on the tennis data,
both calibrated to a common marginal exceedance rate. \vcheck{script 04 plus the Elo comparison at the foot of that script}}
\label{tab:plugin}
\resizebox{\textwidth}{!}{%
\begin{tabular}{lcccc}
\toprule
& \multicolumn{2}{c}{$\bar N=1$} & \multicolumn{2}{c}{$\bar N=2$} \\
\cmidrule(lr){2-3}\cmidrule(lr){4-5}
Functional & as observed & posterior & as observed & posterior \\
\midrule
Longest run of $N_t\geq3$ & 39 & 52 (19, 112) & 359 & 293 (178, 420) \\
Total periods $N_t\geq3$  & 127 & 156 (80, 249) & 815 & 784 (683, 872) \\
\bottomrule
\end{tabular}}
 
\end{table}
 
Section~3.4 of the main paper established how the functionals behave when the
truth is
known. It remains to ask what difference propagation makes on the tennis data
themselves. Table~\ref{tab:plugin} compares the two. Once the marginal rate is
held fixed, the value obtained by treating ratings as observed falls inside
the posterior interval for both functionals at both thresholds, and at the
looser threshold the two agree within four per cent on the rate functional. On
these data, therefore, treating ratings as observed is adequate.
 
This could not have been established in advance. Section~3.4 of the main paper
shows that the longest run is liable to twofold inflation, and nothing in the data indicates which regime obtains. Propagation did not overturn a conclusion here; it established that one could be drawn.

\end{document}